\documentclass{aa}
\usepackage{graphicx}

\newcommand{\kms}       {km~s$^{-1}$}
\newcommand{\h}         {$h^{-1}_{75}\,$~kpc}
\newcommand{\etal}      {{et~al.}}

\newcommand{\pcm}       {cm$^{-2}$}
\newcommand{\lya}       {Ly$\alpha$}

\begin{document}
   \title{21-cm H~I emission from the Damped Lyman-$\alpha$ absorber SBS~1543+593}


   \author{David V. Bowen\inst{1}
          \and
	  Walter Huchtmeier\inst{2}
	  \and
	  Elias Brinks\inst{3}
	  \and
          Todd M. Tripp\inst{1}
	  \and
	  Edward B. Jenkins\inst{1}
          }

   \offprints{D. V. Bowen, e-mail: {\bf{dvb@astro.princeton.edu}}}

   \institute{Princeton University Observatory, Princeton, NJ 08544, USA
         \and
	     Max-Plank-Institut f\"{u}r Radioastronomie, Auf dem
              H\"{u}gel 69, 53121 Bonn, Germany
	\and
	     Depto. de Astronom\'{\i}a, Apdo. Postal 144, Guanajuato,
              Gto. 36000, Mexico
             }

   \date{Received March 12, 2001; accepted April 24, 2001}

   \abstract{
We detect 21-cm emission from the Low Surface Brightness (LSB) galaxy
SBS~1543+593, which gives rise to a Damped-\lya\ (DLy$\alpha$)
absorption line in
the spectrum of the background QSO HS~1543+5921 ($z_e\:=\:0.807$). We
obtain an accurate measure of the velocity of the H~I gas in the LSB
galaxy, $v\:=\:2868$~\kms, and derive a mass of
$M_{\rm{H\,I}}\:=\:1.3\times10^9\:M_\odot$. We compare $M_{\rm{H\,I}}$
with limits obtained towards two other $z\:\sim\:0.1$ D\lya\
systems, and show that SBS~1543+593 would not have been detected. Hence LSB
galaxies similar to SBS~1543+593 can be responsible for D\lya\ systems
at even modest redshifts without being detectable from their 21-cm
emission.
   }

   \maketitle
%

\section{Introduction}

In a recent paper, \cite{Kanekar} reported an attempt to detect 21-cm
emission from the candidate $z\:=\:0.101$ Damped Ly$\alpha$ (D\lya)
absorber towards PKS~0349$-$433.  At these low redshifts, such a
search is entirely warranted, since normal gas-rich late-type galaxies
should be detectable from their 21-cm emission if responsible for the
absorption. As the authors themselves pointed out, such observations
are necessary to establish whether D\lya\ systems have anomalously
high H~I content, irrespective of their optical luminosity or physical
morphology. Significantly, Kanekar~\etal\ failed to detect any
emission from the candidate D\lya\ system. Since their detection
limits were sufficient to find any normal high-mass spiral galaxy,
they concluded that a low-mass galaxy might well be responsible for
the absorber.

Such a conclusion is consistent with results from
ground-based and Hubble Space Telescope images of fields around QSOs
known to show $z<1$ D\lya\ systems. A wide variety of galaxy types
have been identified as possibly responsible for the absorption,
including normal early- and late-type high surface brightness (HSB)
spirals, and amorphous low surface brightness (LSB) galaxies
(Steidel~\etal\ 1991; Steidel~\etal\ 1993; Lanzetta~\etal\ 1997; Le
Brun~\etal\ 1997; Rao \& Turnshek 1998; Pettini~\etal\ 2000;
Turnshek~\etal\ 2001; Cohen 2001).  This wide variety of absorber
types has led to the suggestion that D\lya\ systems may not arise from
normal gas-rich HSB spiral galaxies alone, as has been assumed since
they were first detected (\cite{wolfe86}).

We recently discovered (Bowen~\etal~2001, hereafter Paper~I) that the
nearby LSB galaxy SBS~1543+593 gives rise to a D\lya\ system in the
spectrum of the background QSO HS~1543+5921 ($z_{e}\:=\:0.807$).  The
line of sight to the QSO passes within 2.4$''$ ($\equiv\:0.3
h_{75}^{-1}$ kpc) from the center of the
galaxy\footnote{$h_{75}\:=\:H_0/75$ km~s$^{-1}$ Mpc$^{-1}$, where $H_0$ is the Hubble
constant, and $q_0\:=\:0.0$ is assumed throughout this paper}, and an
H~I column density of $N$(H~I)$=20.35$ dex is measured along the
sightline to the quasar. The galaxy has a central surface brightness
of $\mu_B(0)\:=23.2$ mag arcsec$^{-2}$ and an absolute magnitude of
$M_R\:=\:-16.5$. The redshift of the galaxy, $z=0.009$, was measured
by \cite{RH} (who discovered the pairing of QSO and galaxy) from an
isolated H~II region in a southern spiral arm; no other estimate of
the galaxy's systemic redshift is available.

In this paper, we present the detection of 21-cm emission from
SBS~1543+593, allowing us to determine more accurately the systemic
velocity of the galaxy, as well as its H~I mass. We compare the latter
quantity with that derived by \cite{Kanekar}, as well as a limit
obtained by \cite{Lane} for a second low-$z$ D\lya\ absorber, and show
that SBS~1543+593 would not have been detected in either survey.  Hence LSB
galaxies similar to SBS~1543+593 can be responsible for D\lya\ systems
at even modest redshifts without being detectable from their 21-cm
emission.


\section{Observations \& Results}

We initially checked to see if HS~1543+5921 was radio-loud, in the
hope of using the background source to search for 21-cm absorption
from the foreground galaxy.  Using the NVSS database (\cite{condon})
we found no coinciding radio emission above a $4 \sigma$ noise level
of about 1.8 mJy. An unrelated 32 mJy flux density source, unresolved
at the resolution of the NVSS survey (typically $45''$), lies $2'$ to
the southeast of the expected position of the QSO.

21-cm line emission observations were made 22nd January 2001, using
the 100 m radio telescope at Effelsberg, which has a half power beam
width of 9.3$'$. The `total power mode' was applied, i.e., a reference
field was observed 5 min earlier in Right Ascension and later
subtracted from the on-source observation in order to reduce
instrumental effects.  15 mins were spent on-source, and 15 mins
off-source. The two channel 21-cm receiver (with a system noise of
30~K) was followed by a 1024 channel autocorrelator split into four
banks of 256 channels each.  A total bandwidth of 6.25 MHz yielded a
channel separation of 5.2~\kms\ and a velocity resolution of 6.3~\kms
. A second order polynomial was applied as a baseline correction.
The rms error was determined to be 2.1~mJy per channel.

\begin{figure}
\centering
\includegraphics[width=9cm]{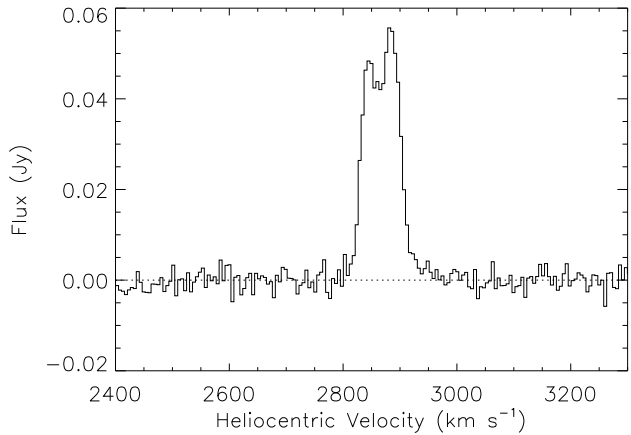}
\caption{21-cm emission from SBS~1543+593. The bulk of the emission
arises at 2868~\kms, with two peaks arising from the rotating,
inclined disk at 2843.5 and 2884.0~\kms .}
\label{Fig:HI}
\end{figure}

The 21-cm emission detected from SBS~1543+539 is shown as a function
of heliocentric velocity in Figure~\ref{Fig:HI}. The center of the
emission measured from the FWHM of the profile is 2868$\pm2$~\kms\
($cz=0.0096$) and
shows the classic double-horn peaks, with maxima at 2843.5 and 2884.0
\kms . Such profile shapes are well understood to arise from the disks
of spiral galaxies with flat rotation curves and exponential gas
distributions (\cite{GH88}). We also note though that the profile
resembles those studied by \cite{matthews} for a sample of extreme
late-type spirals, whose single-dish emission line profiles are
``filled in'' between the rotation peaks, perhaps due to the rotation
curves of the galaxies rising more slowly than those of typical spiral
galaxies.  SBS~1543+539 may have a similar characteristic, which can
be tested with more extensive two-dimensional 21-cm H~I mapping. The
velocity of the 21-cm emission agrees reasonably well with the value
of 2700~\kms\ (with no quoted error) measured by Reimers \& Hagen
(1998). Their value was derived, essentially, from measuring the
velocity of H$\alpha$ and [O~III] emission lines in low
signal-to-noise data with a resolution of $\sim 18$~\AA, or $\sim
800-1300$~\kms .

Comparing the redshift of SBS~1543+593 with the metal absorption lines
detected in Paper~I is not so straightforward. The HST spectra in
which the lines were observed were recorded with the STIS G140L
grating and the 52x0.5 aperture, for which there is considerable
uncertainty in the zero point of the wavelength calibration. This can
in principle be corrected by comparing the velocities of Galactic
low-ionization metal lines with the velocity of H~I emission along the
line of sight. Unfortunately, the low resolution of the G140L
(200$-$300~\kms ) and the low signal-to-noise of the data make the
precise velocity of the Galactic absorption hard to
determine. Further, inspection of the 21-cm emission along the
sightline [taken from the Leiden/Dwingeloo 21-cm survey (\cite{hart})]
shows that besides the bulk of the emission around 0~\kms, there are
strong peaks from Galactic High Velocity Clouds at $v\:\sim\:-140$ and
$-70$~\kms . Our final estimate of the absorber's velocity is
$\sim3050$~\kms , apparently in poor agreement with the 21-cm
emission. Nevertheless, the error in this number is large,
$\sim\:100-200$~\kms, and higher resolution UV observations are
clearly needed to better tie down the redshift of the absorbing gas
before considering a real discrepancy between the bulk of the 21-cm
gas and the absorbing material.


We calculate the total mass of the H~I gas in the usual manner:

\[
M_{\rm{H\,I}}\:=\:2.36\times 10^5
D({\rm{Mpc}})^2\:\int\,S({\rm{Jy}})\,dv({\rm{km\,s^{-1}}})\:\:M_\odot\: .
\]

\noindent
where $D$ is the luminosity distance to the galaxy, taken to be 38~Mpc 
for the adopted cosmology, and $S$ is the 21-cm
flux, which is integrated over the observed velocity range of the
profile. For SBS~1543+593 we derive a flux of 3.9~Jy~\kms\ over a
velocity range of 2740$-$3000~\kms , which gives

\[
M_{\rm{H\,I}}\:=\:1.3\times10^9\: h_{75}^{-2}\:  \:M_\odot\:.
\]

The statistical error on the flux is 0.048~Jy~\kms , but
this is likely an underestimate of the true error, since
systematic errors also contribute. Experience with similar
datasets suggests an error of $\sim 10$~\% is more realistic, which
would lead to an uncertainty of $\sim 1\times 10^8~M_\odot$ in the
H~I mass of SBS~1543+593.

The width of the H~I profile displayed in Fig. 1 corresponds to a lower
limit to the dynamical mass of

\[
M_{\rm{dyn}}\: = \: 0.76 \times 2.33 \times 10^5 (\delta V)^2 \: R\:M_\odot\:
\]

\noindent
where $\delta V$ is the half width at zero intensity and $R$ the radius
of the galaxy (see, e.g., \cite{vanM}). In Paper~I we were able to
trace the $R$-band light of SBS~1543+593 out to a radius of $\sim
30''$, or 4.1~\h . Taking this value to be $R$, and using $\delta V =
62$~\kms , we find:

\[
M_{\rm{dyn}}\: \geq 2.8\times10^{9}\: h_{75}^{-1} \:M_\odot\:.
\]

\noindent
This is a lower limit because the 
rotational velocity has not been corrected for the (unknown) inclination.
Although measuring the inclination is difficult
(see Paper~I), the galaxy is clearly far from being edge-on, so
$M_{\rm{dyn}}$ is a conservative limit. Nevertheless, the dynamical
mass is at least of order 50\% higher than the gas mass.

In Paper~I we estimated the $R$-band magnitude of the galaxy to be
$R\:=\:16.3$. We can correct to a $B$-band magnitude by adopting the
correction found by de Blok~\etal\ (1996) for their sample of LSB
galaxies, and derive $B\:=\:R+0.78\:=\:17.1$. This allows us to
calculate the usual mass-to-light ratio of the galaxy, $M/L$, where
$L$ is derived from the $B$-band luminosity. Since the absolute
magnitude of SBS~1543+593 is $M_B\:=\:-15.8$, 
then its total luminosity is
$L/L_\odot\:\sim\:3\times10^8$, which gives 

\[
M_{\rm{H\,I}}/L\:\sim\:4 .
\]

In this case, the error in calculating $M_{\rm{H\,I}}/L$ is likely dominated
by the value of $L$ rather than $M_{\rm{H\,I}}$, since contamination of
the galaxy's light from the QSO and close-by star may cause $L$ to be
overestimated (see Paper~I).

\section{Conclusions}

The H~I mass derived for SBS~1543+593 is about 1/5 that of the Milky
Way, and is close to the median value of $M_{\rm{H\,I}}$ found by
\cite{deblok} for a sample of 19 late-type LSB galaxies.  Assuming
$R-I\:\approx\:0$, it appears that the LSB galaxy is gas-rich compared
to other LSB galaxies studied by de Blok~\etal , although there are
few galaxies at such low total luminosities in their sample. The value
of $M_{\rm{H\,I}}/L\:\sim\:4$ for SBS~1543+593 is correspondingly
high, since the typical value for galaxies in de Blok~\etal 's sample
was $\leq 1$. On the other hand, the 38 LSB galaxies studied by
Karachentsev~\etal\ (2001, and refs. therein) in the magnitude range
$-16.2 < M_B < -15.1$ have an average value of
$M_{\rm{H\,I}}/L\:=\:2.6$ with a scatter between 0.3 and 11.8. The value
for SBS~1543+593 is obviously well within this range.

With the derived $M_{\rm{H\,I}}$, we can consider how easily
SBS~1543+593 would be identified as the origin of a D\lya\ system if
at higher redshift.  It is complicated to calculate precisely whether
a particular set of {\it optical} observations would reveal
SBS~1543+593 much beyond its redshift of 0.0096. For example, the
success in adequately subtracting the QSO profile to reveal overlying
galaxies depends on the signal-to-noise of the data, and the size and
stability of the seeing during the observations. Similarly, an image's
ability to record LSB features depends on the detector's physical
characteristics (particularly its spatial scale), and the proximity of
the galaxy to the QSO itself. Observing the 21-cm emission line
however, does not suffer from these difficulties, particularly if the
background QSO is not radio-loud, and in principle, picking out the
emission at the redshift of the D\lya\ system should be a relatively
`clean' observation. The problem, of course, is that the flux from the
low surface brightness of the H~I emission can only be detected if the
galaxy is at a low redshift. Hence searches for 21-cm emission from
D\lya\ absorbers are confined to the very lowest redshift systems
known.

Kanekar~et~al.~(2001) failed to detect 21-cm emission from the
$z\:=\:0.101$ candidate D\lya\ absorber towards PKS~0439$-$433. The
absorber is a ``candidate'' because the \lya\ absorption line has not
yet been observed in the ultraviolet, although the strength of
low-ionization lines in the system suggest the H~I column density
should be high, $N$(H~I)$\sim10^{20}$~\pcm\
(\cite{petitjean}). Kanekar~\etal\ set a $3\sigma$ limit of
$M_{\rm{H\,I}}\:<\:1.8\times 10^9\: h_{75}^{-2} \:M_\odot$ for the system assuming a
velocity spread equal to their resolution, 30~\kms; such a limit would
be insufficient to detect SBS~1543+593, and hence would not rule out
such an LSB galaxy as being responsible for the $z\:=\:0.101$
system. \cite{Lane} also searched for 21-cm
emission from the 21-cm absorber at $z\:=\:0.0912$ towards B~0738+313,
and again failed to detect any flux. Their observations were slightly
more sensitive, reaching a $3\sigma$ limit of
$M_{\rm{H\,I}}\:<\:6.5\times 10^8\: h_{75}^{-2} \:M_\odot$ assuming a velocity
resolution of 22.5~\kms . Even here, however, if we re-calculate
$M_{\rm{H\,I}}$ by instead considering the velocity range over which
we detect 21-cm emission from SBS~1543+593, some 124~\kms , this limit
would increase to $M_{\rm{H\,I}}\:<\:3.8\times 10^9\: h_{75}^{-2} \:M_\odot$. Again,
this would be too high a limit to permit detection of SBS~1543+593 at
$z\:=\:0.0912$.  Using this velocity range for Kanekar~et~al.'s
observations would  increase their limit to $M_{\rm{H\,I}}$
from the absorber towards PKS~0439$-$433 as well.

It is relatively straightforward to calculate the redshift out to
which a galaxy like SBS~1543+593 {\it could} be detected given our
observational set-up. We would measure a $3\sigma$ flux of 1~Jy~\kms\ over 20
channels from a galaxy with $M_{\rm{H\,I}}\:=\:1.3\times10^9\:
h_{75}^{-2}\: \:M_\odot$ at a distance of 94~Mpc, or $z=0.023$ for
$H_0=75$~\kms~Mpc$^{-1}$. In actuality, we could extend this limit by
observing at lower resolution: rebinning by a factor of 4 (20~\kms\
per channel) and assuming that the error in the flux would be reduced
by a factor of two would enable us to detect SBS~1543+593 at the
3$\sigma$ level at a distance of 136~Mpc, or $z=0.034$.

As noted in Paper~I, the detection of a D\lya\
system from SBS~1543+593 does not prove that {\it all} D\lya\ systems
arise in LSB galaxies. Nevertheless, as evidence accumulates to show
that at least some D\lya\ absorbers must be dwarf or LSB
galaxies, our data demonstrates that the non-detection of 21-cm
emission towards higher redshift systems is entirely consistent with
the H~I mass of one particular LSB galaxy---SBS~1543+593---known
unequivocally to be a D\lya\ system.

\begin{acknowledgements}
We thank the referee, Veronique Cayatte for helpful comments. Support
for this work was provided in part from NASA through grant NAS5-30110.

\end{acknowledgements}

\end{document}